\documentclass[twocolumn]{aastex701}
\usepackage{threeparttable}

\begin{document}

\title{Spectral study of the outburst decay of the accreting millisecond X-ray pulsar SRGA J144459.2-604207}

\author[orcid=0000-0001-8345-3125,sname=Chandra]{Amar Deo Chandra}
%\altaffiliation{Kitt Peak National Observator}
\affiliation{Aryabhatta Research Institute of Observational Sciences, Manora Peak, Nainital, Uttarakhand, 263001, India}
%\correspondingauthor{A.D.~Chandra}
\email[show]{amar.deo.chandra@gmail.com}
\begin{abstract}
We study the spectra of the accreting X-ray millisecond pulsar SRGA J144459.2-604207 during the 2024 outburst using the Neutron star Interior Composition Explorer (NICER) and Swift observations. The spectra during the outburst decay, reflares, and quiescent state are explored using the absorbed Comptonized model. We find that the spectra during the quiescent state can also be explained using the absorbed power-law and absorbed blackbody model. The spectral evolution of the source is explored as the outburst decays into quiescence.
We study the long-term quiescent X-ray activity of the source spanning roughly 45 years and find that the long-term quiescent luminosity may be explained using the deep crustal heating model. We also find that the coronal activity of the companion star alone cannot power the quiescent X-ray luminosity of the source. We place the source on the radio-X-ray luminosity plane and compare its position with other sources. We estimate the propeller luminosity of the source and find that it is smaller than the estimated luminosity during reflares and the quiescent state during the 2024 outburst.
Several reflares are detected during the outburst decay, one of which is near-simultaneous with the detection of an ultrafast outflow and radio emission. We explore plausible mechanisms that may power outflow, radio emission and associated jet formation in this accreting binary.
\end{abstract}



\section{Introduction}
Accreting millisecond X-ray pulsars (AMXPs) consist of a fast-rotating neutron star (P$\lesssim$10 ms) and a low-mass companion ($\lesssim 1 M_{\odot}$). They belong to the category of Low Mass X-ray Binary (LMXB) systems and have typical magnetic fields of $\sim 10^{8-9}$G \citep{mukherjee2015magnetic}, which are roughly 2-3 magnitudes lower than the magnetic fields detected in High Mass X-ray Binary (HMXB) systems \citep{coburn2002magnetic,chandra2020study,chandra2023astrosat}. Accretion onto the neutron star in LMXB systems occurs via Roche lobe overflow from the companion star \citep{frank2002accretion}, unlike HMXBs, which accrete from the stellar wind \citep{tauris2006formation,chandra2021detection}. AMXPs are known to usually be in the quiescent state, having a low luminosity of $\sim 10^{31-33}$\,erg\,s$^{-1}$, but exhibit sporadic X-ray outbursts having luminosity $\sim 10^{36-37}$\,erg\,s$^{-1}$ \citep{patruno2020accreting,salvo2021accretion}.

SRGA J144459.2-604207 was discovered on 2024 February 21 during an ongoing all-sky survey by the Mikhail Pavlinsky Astronomical Roentgen Telescope X-ray Concentrator (ART-XC) telescope onboard the Spectrum-Roentgen-Gamma (SRG) observatory \citep{mereminskiy2024srg}. The transient was also detected using follow-up Swift/XRT observations \citep{chandra2024swift}. Coherent X-ray pulsations were discovered at $\sim$447.9 Hz using follow-up observations with the Neutron star Interior Composition Explorer (NICER), which confirmed that the new source was an AMXP \citep{ng2024nicer}. In addition, a type I X-ray burst was also detected during initial NICER follow-up observations \citep{ng2024nicer1,ng2024nicer}. Observations using the Monitor of All-sky X-ray Image (MAXI) suggested that the source had been brightening since 2024 February 15 \citep{mihara2024maxi}. Radio
emission from the source was detected by the Australia Telescope Compact Array (ATCA) on 2024 February 29 at 5.5 GHz and 9 GHz \citep{russell2024atca} using the refined Chandra location of the source \citep{illiano2024chandra}. No optical/IR counterpart was detected during optical/IR observations of the transient \citep{sokolovsky2024no,cowie2024srga,saikia2024optical,saikia2024search}. Regular Type-I X-ray bursts were detected from the source using ART-XC, NICER, NinjaSat, Nuclear Spectroscopic Telescope Array (NuStar), Insight-Hard X-ray Modulation Telescope (HXMT), Imaging X-ray Polarimetry Explorer (IXPE), and X-ray Multi-Mirror Mission (XMM)-Newton observations \citep{papitto2025discovery,molkov2024discovery,takeda2025ninjasat,sanchez2024integral,sanchez2024srga,fu2025comprehensive,malacaria2025disk,li2025timing}. The orbital period of the system was inferred to be $\sim$5.2 hr from the sinusoidal Doppler shift of the spin frequency obtained from NICER and Insight-HXMT observations, and the estimated lower limit on the mass of the companion was pegged at $\sim$0.255 $M_{\odot}$ \citep{ng2024nicer,li2025timing}.
The distance to the source was estimated to be $\sim$10 kpc obtained from Photospheric Radius Expansion (PRE) bursts detected by the Insight-HXMT \citep{fu2025comprehensive}. Polarized emission was detected from the source having an average polarization
degree of 2.3\% ± 0.4\% using IXPE observations \citep{papitto2025discovery}. The source was found to be in an active state during 2022 January and 2023 December from the archival MAXI and International Gamma-ray Astrophysics Laboratory (INTEGRAL) observations \citep{negoro2024maxi,sguera2024integral}.

In this paper, we investigate the spectral evolution of SRGA J144459.2-604207 during the decay of the 2024 outburst as the source becomes quiescent using NICER and Swift observations. The paper is organized as follows. We describe observations from the NICER and the Swift observatories and their data analysis procedures in Section 2. In Section 3, we carry out spectral studies of the source using NICER and Swift observations during the 2024 outburst decay and quiescence. We also study the long-term quiescent X-ray activity of the source spanning roughly 45 years from archival satellite observations.
In Section 4, we discuss possible mechanisms that can power X-ray emission from the source during the decay of the X-ray outburst and as the pulsar becomes quiescent.
 We explore the radio-X-ray luminosity relation of the source and discuss putative mechanisms which can power near-simultaneous outflow, radio emission, and possible jet formation in the accreting binary. We summarise our findings in Section 5.

\section{Observations and data reduction}
The one-day averaged monitoring observations of SRGA J144459.2-604207 from the Monitor of All-sky X-ray Image (MAXI, \citep{matsuoka2009maxi}) mission\footnote{\url{https://maxi.riken.jp/star_data/J1444-607/J1444-607.html}\label{fn:note1}} in the 2-20 keV energy band during the period MJD 60330 (2024 January 21) until MJD 60450 (2024 May 20) is shown in Figure \ref{f1}. It is clearly seen from Figure \ref{f1} that the X-ray outburst in this source began around MJD 60355, attained its peak around MJD 60365 and then the outburst gradually decayed until around MJD 60380, after which the source entered the quiescent regime. However, a few possible reflares are visible in the MAXI light curve around MJD 60387 and MJD 60424. We analyse Swift observations of SRGA J144459.2-604207 during the outburst decay from 2024 March, and these epochs are marked by dashed vertical lines in Figure \ref{f1}. The source was observed on multiple occasions using the Neutron star Interior Composition Explorer (NICER, \citet{gendreau2016neutron}) telescope from 2024 February 21 (MJD 60361.83) till 2024 May 3 (MJD 60433.86). The X-ray properties of the source when the source was bright until mid-2024 March was explored in several studies \citep{ng2024nicer,papitto2025discovery,li2025timing}.
We analyse NICER observations from 2024 March 14 until 2024 May 3 to study the spectral evolution of the pulsar, and these epochs are marked by dotted vertical lines in Figure \ref{f1}.

\begin{table*}
\centering
\caption{Observations analyzed in this paper. The last column mentions the state of the source during these observations.
\label{tab1}}
\begin{tabular}{lcccccc}
\hline\hline
Observatory & Obs. ID & Start date (MJD) & End Date (MJD) & Exposure (s) & Comment\\
            \hline
NICER& 	6639080113	& 	60383.104	& 	60383.113	& 	765	& reflare\\
& 	6639080116	& 	60392.050	& 	60392.11	& 	5180	& quiescent\\
& 	6639080119	& 	60397.230	& 	60397.248	& 	1522	& quiescent\\
& 	6639080120	& 	60398.070	& 	60398.075	& 	412	& quiescent\\
& 	6639080121	& 	60409.869	& 	60409.871	& 	133	& quiescent\\
& 	6639080122	& 	60409.998	& 	60410.003	& 	440	& quiescent\\
& 	6639080124	& 	60412.443	& 	60412.45	& 	578	& quiescent\\
& 	6639080126	& 	60416.765	& 	60416.767	& 	139	& reflare\\
& 	6639080133	& 	60424.437	& 	60424.462	& 	2180	& quiescent\\
& 	6639080134	& 	60425.019	& 	60425.04	& 	1847	& quiescent\\
& 	6639080135	& 	60426.831	& 	60426.836	& 	442	& quiescent\\
& 	6639080136	& 	60427.599	& 	60427.608	& 	747	& reflare\\
& 	6639080139	& 	60430.826	& 	60430.83	& 	374	& quiescent\\
& 	6639080141	& 	60433.860	& 	60433.865	& 	446	& quiescent\\
Swift & 00016537004 &  60371.394	& 60371.405 & 924 & outburst decay\\
 & 00016537005 &  60378.313	& 60378.325 & 1070 & outburst decay\\

\hline

\end{tabular}
\end{table*}

\subsection{NICER observations}

The log of NICER observations used in this study is shown in Table \ref{tab1}.
We reduced the data using \textsc{heasoft v6.35.2}\footnote{\url{https://heasarc.gsfc.nasa.gov/docs/software/lheasoft/}} and the NICER Data Analysis Software (\textsc{NICERDAS v14}) using calibration version \texttt{xti20240206}. NICER experienced a light leak starting on 22 May 2023, and it is recommended by the NICER instrument team to use \textsc{heasoft v6.35.2} for processing NICER data\footnote{\url{https://heasarc.gsfc.nasa.gov/docs/nicer/analysis_threads/light-leak-overview/}}, which includes updates for improved screening for all data taken during orbit day since the optical light leak.
We followed the standard data analysis to extract the cleaned event files using \texttt{nicer-l2}.
We extracted the source spectra, the background spectra, the arf, and the response files using the tool \texttt{nicerl3-spect}. We extracted the background spectra using the 3C50 model. For spectral analysis, the spectra were optimally rebinned so that each bin had a minimum
of 25 counts \citep{kaastra2016optimal} using the tool {\tt ftgrouppha}. The source and the background light curves were extracted in the energy band of 0.5-8 keV using the tool \texttt{nicerl3-lc}. The energy range for extracting light curves was limited to 0.5-8 keV, as the observations were dominated by background beyond this energy band.

\subsection{Swift observations}
The Neil Gehrels Swift Observatory \citep{gehrels2004swift} observed
SRGA J144459.2-604207 for six pointings (ObsIds 00016537001-00016537005, 00016537009, Table \ref{tab1}) spanning the period from 2024
February 22 (MJD 60362.1) until 2024 April 6 (MJD 60406.6). Only two Swift/X-Ray Telescope (XRT) observations (ObsIds 00016537004 and 00016537005) out of the six observations were taken during the outburst decay and data from these two observations have been analysed in this paper.
Swift/XRT observations were carried out in the Windowed Timing (WT) mode. The Swift/XRT spectra for each epoch were extracted using the online tools \citep{evans2009methods}\footnote{\url{https://www.swift.ac.uk/user_objects/}}  hosted by the UK Swift Science Data Centre. The on-source exposures are
$\sim$1 ks for each Swift/XRT pointing. The spectra were binned with \texttt{grppha} to have a minimum of 20 counts per bin for the application of $\chi^{2}$ statistics.

\begin{figure}
\centering
  \includegraphics[width=\columnwidth]{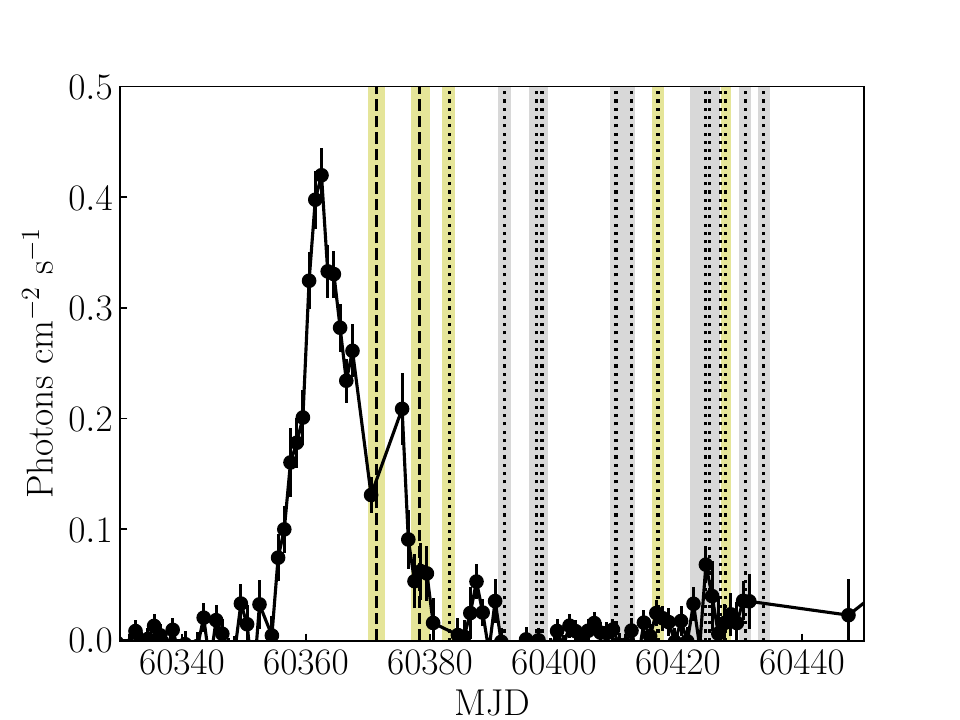}
  \caption{MAXI one-day averaged light curve of SRGA J144459.2-604207 in
the 2-20 keV energy band\footref{fn:note1}
spanning the duration MJD 60330 (2024 January 21) until MJD 60450 (2024 May 20). The dashed and dotted vertical lines indicate the epochs of Swift and NICER observations, respectively. The shaded yellow and gray regions denote the epochs during outburst/reflare and quiescence, respectively.}
 \label{f1}
\end{figure}

\section{Analysis and results}
\subsection{X-ray light curve}
\begin{figure}
\centering
  \includegraphics[width=\columnwidth]{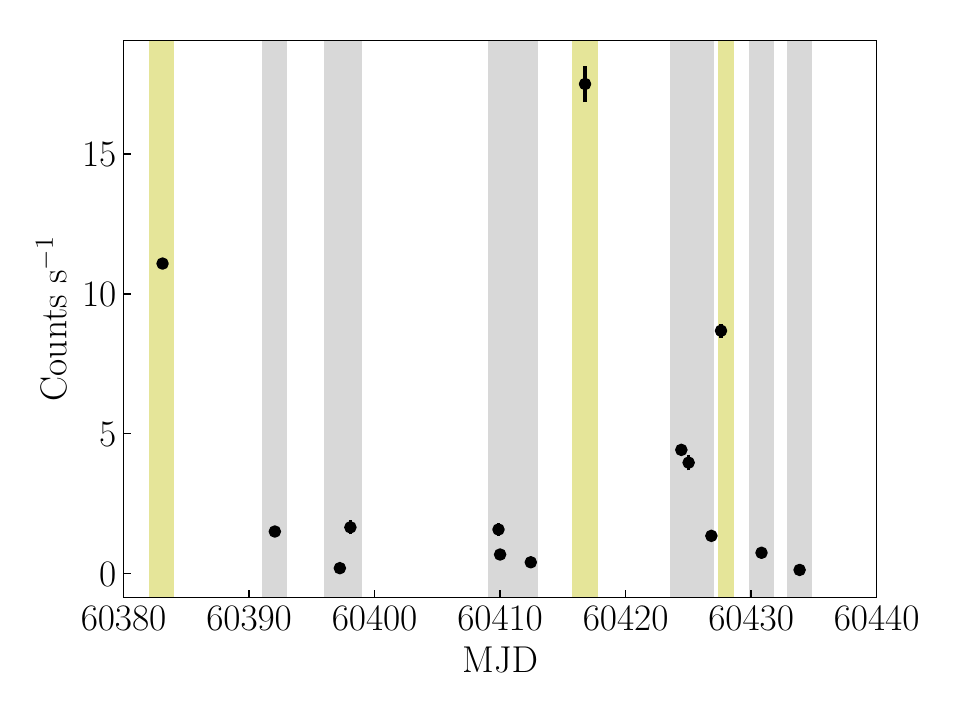}
        \caption{NICER background-subtracted averaged light curve of SRGA J144459.2-604207 in the 0.5-8 keV energy band. The shaded yellow and gray regions denote the epochs during reflare and quiescence, respectively.}
        \label{f2}
\end{figure}

The NICER background-subtracted averaged light curve (0.5-8 keV energy band) of SRGA J144459.2-604207 is shown in Figure \ref{f2}.
The NICER count rates have been averaged for a given epoch. It should be noted that for some epochs, the source count rate was background-dominated, and these were excluded from analysis. It is observed that the NICER count rate during this period varied from $\sim$0.13$~\rm{counts ~s^{-1}}$ to $\sim$17.5 $\rm{counts ~s^{-1}}$, showing a dynamic range of $\sim$130. A count rate threshold of $\sim$5$~\rm{counts ~s^{-1}}$ (approximately three times the average count rate during quiescence) was used for classifying a reflare in the NICER light curve. The NICER count rate gradually decreased from $\sim$11$~\rm{counts ~s^{-1}}$ (MJD 60383.1) to $\sim$1$~\rm{counts ~s^{-1}}$ (around MJD 60400) and remained at this level until around MJD 60412. The NICER count rate jumped to $\sim$17.5$~\rm{counts ~s^{-1}}$ around MJD 60416.8, suggesting reflaring state of the source. Thereafter, the count rate gradually decreased to $\sim$0.1$~\rm{counts ~s^{-1}}$ in roughly a couple of weeks, suggesting that the source had entered the quiescent state. The MAXI one-day averaged light curve of the source also showed an increase in X-ray activity around MJD 60420. The dynamic range during this period after the reflaring state of the source was $\sim$125, which is similar to that estimated earlier from NICER observations.

\subsection{Spectral analysis}

\begin{figure*}
\plottwo{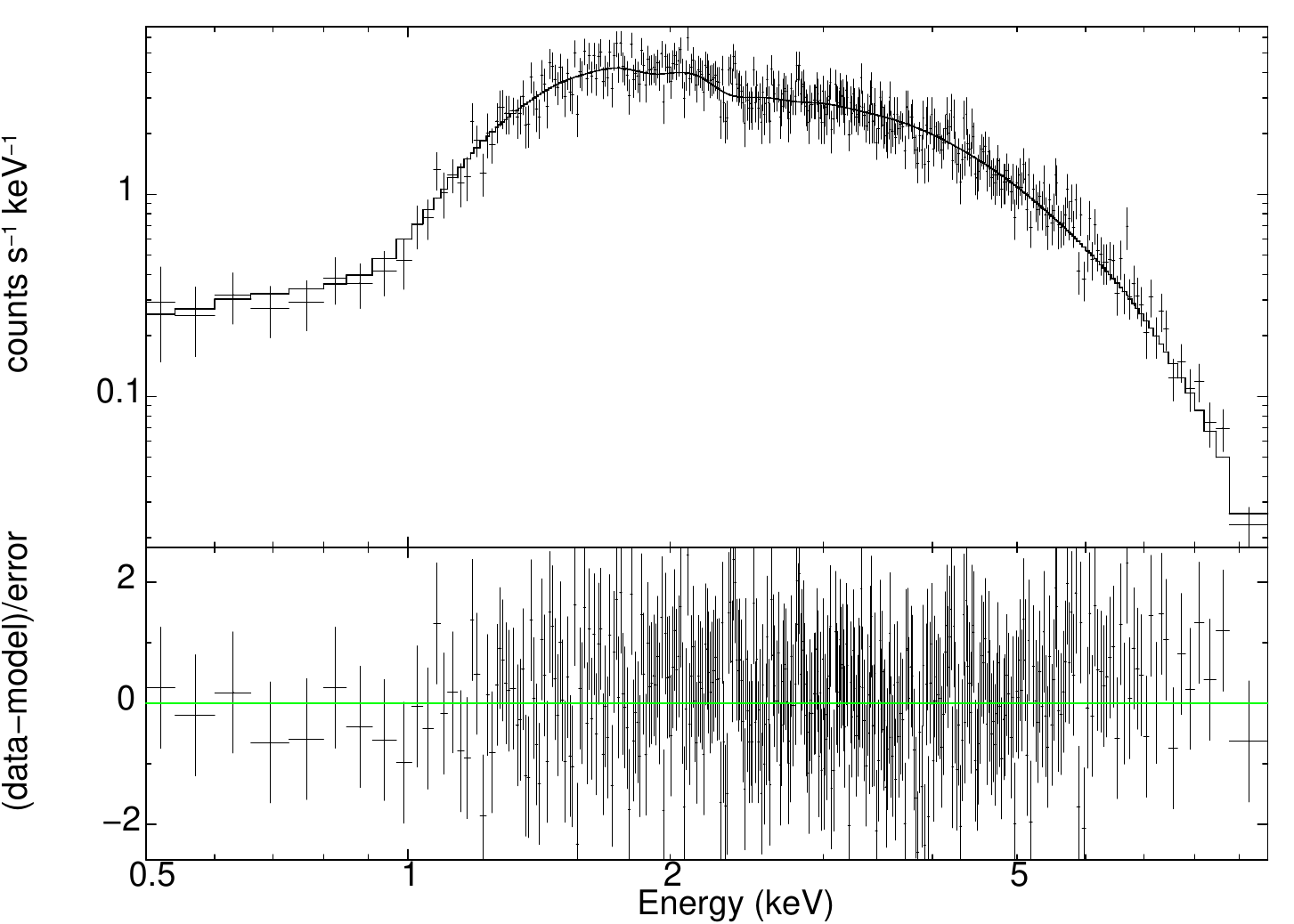}{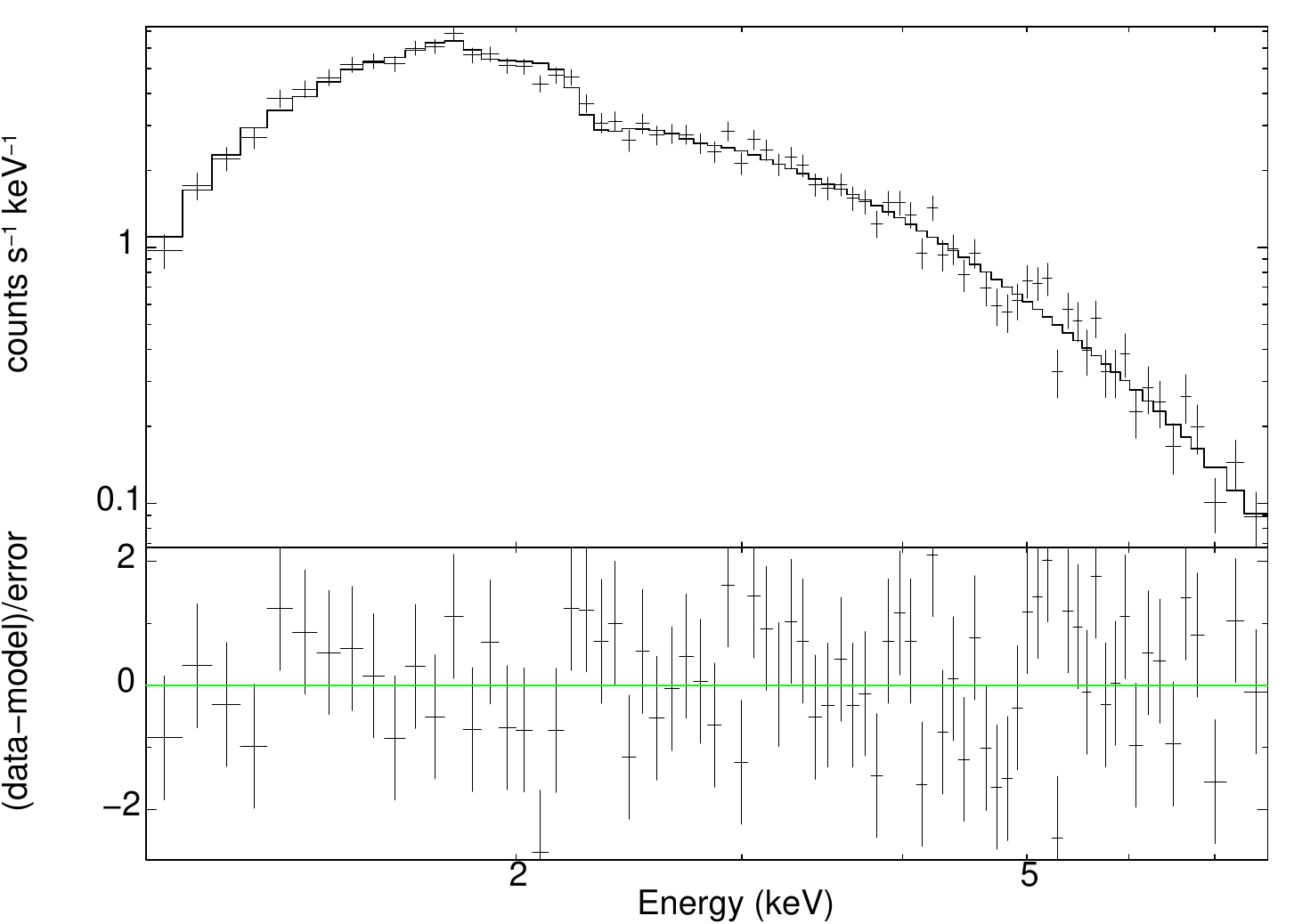}
        \caption{Left: Swift/XRT spectra (during outburst
decay) for 2024 March 2 (MJD 60371.4) fitted with the absorbed Comptonized model. The residuals between the data and the model are shown in the lower panel. Right: NICER spectra (during reflare) for 2024 March 14 (MJD 60383.1) fitted using the absorbed Comptonized model.}
        \label{f3}
\end{figure*}

\subsubsection{Spectra during outburst decay}
The Swift/XRT spectra (Obs. IDs 00016537004 and 00016537005) for each epoch were fitted separately using \textsc{xspec} 12.15.0d \citep{arnaud1996astronomical}.
We used the thermally Comptonized
continuum model {\tt nthcomp} \citep{zdziarski1996broad,zycki1999} modified by the interstellar absorption ({\tt tbabs$\times$(nthcomp)}) to fit the spectra. The {\tt nthcomp} model is parameterized by an asymptotic
power-law photon index $\Gamma$, temperatures of the
electron cloud ($kT_e$), and seed photons ($kT_{bb}$). The seed photons were assumed to be blackbody seed photons emanating from the surface of the neutron star. We have used the {\tt{tbabs}} model \citep{wilms2000absorption} for absorption in the spectrum during spectral fitting. The photoelectric
absorption cross-section used in the spectral fitting was {\tt vern} \citep{verner1996atomic}. All uncertainties of the spectral parameters are reported at a 1$\sigma$ confidence level for a given parameter. The results of our spectral analysis during outburst decay are summarized in Table \ref{table:spec_1}.

Figure \ref{f3} shows the fitted Swift/XRT spectra for 2024 March 2 (MJD 60371.4). The spectra were fitted in the energy range of 0.5-10 keV using the absorbed thermal Comptonized model. The fitted hydrogen column density was $N_{\rm H} = (2.9\pm0.3)\times 10^{22}$ cm$^{-2}$. The fitted $N_{\rm H}$ is more than the interstellar absorption ($N_{\rm H} \sim 1.7 \times 10^{22}$ cm$^{-2}$) along the direction of this source \citep{bekhti2016hi4pi} \footnote{\url{https://heasarc.gsfc.nasa.gov/cgi-bin/Tools/w3nh/w3nh.pl}}.
We obtained photon-index $\Gamma$ = 2.01$\pm$0.06, with $\chi^{2}$ = 336.83, d.o.f. = 378, and reduced $\chi^{2}$ = 0.9 (Table \ref{table:spec_1}). The electron temperature ($kT_e$) could not be constrained and was fixed at 5 keV, which is similar to that obtained from the NICER spectral fitting \citep{li2025timing} near this epoch. The temperature of the blackbody seed photons ($kT_{bb}$) obtained was $0.27\pm0.10$ keV. The flux was computed using the {\tt cflux} model, and the model was extrapolated out to 0.5-100 keV by using the {\tt energies} command in \textsc{xspec}. The unabsorbed flux in the energy range 0.5-10 keV
 is $1.5\pm0.3 \times10^{-9}$ erg cm$^{-2}$ s$^{-1}$, which implies a source luminosity of $1.8\pm0.4 \times10^{37}$ erg s$^{-1}$ for a distance of $\sim$10 kpc. The estimated luminosity in the 0.5-100 keV energy range is $1.9\pm0.4 \times10^{37}$ erg s$^{-1}$. The spectral parameters for the fitted Swift/XRT spectra for 2024 March 9 (MJD 60378.3) are shown in Table \ref{table:spec_1}.

 \begin{table*}[hbtp]
\begin{ruledtabular}
\centering
 \caption{\label{table:spec_1} Best-fit spectral parameters of the Swift and NICER data for SRGA J144459.2-60420 during the outburst decay and reflare, respectively, using the model  {\tt tbabs$\times$(nthcomp)}.}
 \centering
 \begin{tabular}{llll}
&Swift  & Swift  & NICER \\
Epoch & MJD 60371.4  &  MJD 60378.3  & MJD 60383.1 \\
 \hline

Parameter (units) &Best-fit values & \\

  \hline
$N_{\rm H}~(10^{22}~ {\rm cm}^{-2})$& $ 2.9\pm0.3 $ & $ 2.1\pm0.5 $ & $  2.46\pm0.54$ \\
$\Gamma$                            & $ 2.01 \pm0.06 $  & $ 1.8 \pm0.1 $    & $  1.94\pm0.14$ \\
$kT_{\rm e}$ (keV)                        &  5 (fixed)  &  5 (fixed)    & $3\pm1$ \\
$kT_{\rm bb}$ (keV)                       & $ 0.27 \pm0.10 $  & $0.30 \pm0.17$   & $0.2\pm0.1$ \\
  \hline
 $\chi^{2}/{\rm d.o.f.}$ &   336.83/378 &   190.15/173& 82.37/71\\
 $F$ ($10^{-9}$ erg s$^{-1}$ cm$^{-2}$)\textsuperscript{a} & $1.5\pm0.3$ & $0.38\pm0.11$ & $0.13\pm0.03$ \\
\end{tabular}
\end{ruledtabular}
\begin{tablenotes}
\item[a]  \textsuperscript{a} Unabsorbed flux in the 0.5--10 keV energy range.
\end{tablenotes}
 \label{tab:broadband_spec}
\end{table*}

\subsubsection{Spectra during reflare}
The NICER spectra (Obs. ID 6639080113) was fitted using the {\tt tbabs$\times$(nthcomp)} model. The NICER spectra for each epoch were fitted separately. The exposure was low for other reflare epochs (Obs. IDs 6639080126 and 6639080136) for spectral analysis. Figure \ref{f3} shows the fitted NICER spectra for 2024 March 14 (MJD 60383.1). The spectral fitting was restricted to the energy band of 1-8 keV as the spectra were dominated by the background beyond this energy range. We obtained hydrogen column density $N_{\rm H} = (2.46\pm0.54)\times 10^{22}$ cm$^{-2}$ and a photon-index $\Gamma$ = 1.94$\pm$0.14, with $\chi^{2}$ = 82.37, d.o.f. = 71, and reduced $\chi^{2}$ = 1.16 (Table \ref{table:spec_1}). The inferred electron temperature ($kT_e$) and the temperature of the blackbody seed photons ($kT_{bb}$) were $3\pm1$ keV and $0.2\pm0.1$ keV, respectively. The unabsorbed flux in the energy range 0.5-10 keV
 is  $1.3\pm0.3 \times10^{-10}$ erg cm$^{-2}$ s$^{-1}$, which implies a source luminosity of $1.6\pm0.4 \times10^{36}$ erg s$^{-1}$ for a distance of $\sim$10 kpc. The estimated luminosity in the 0.5-100 keV energy range is $1.7\pm0.4 \times10^{36}$ erg s$^{-1}$. The source luminosity dropped by roughly an order of magnitude compared to that during 2024 March 2.

\subsubsection{Spectra during quiescent state}
\begin{figure*}
\plottwo{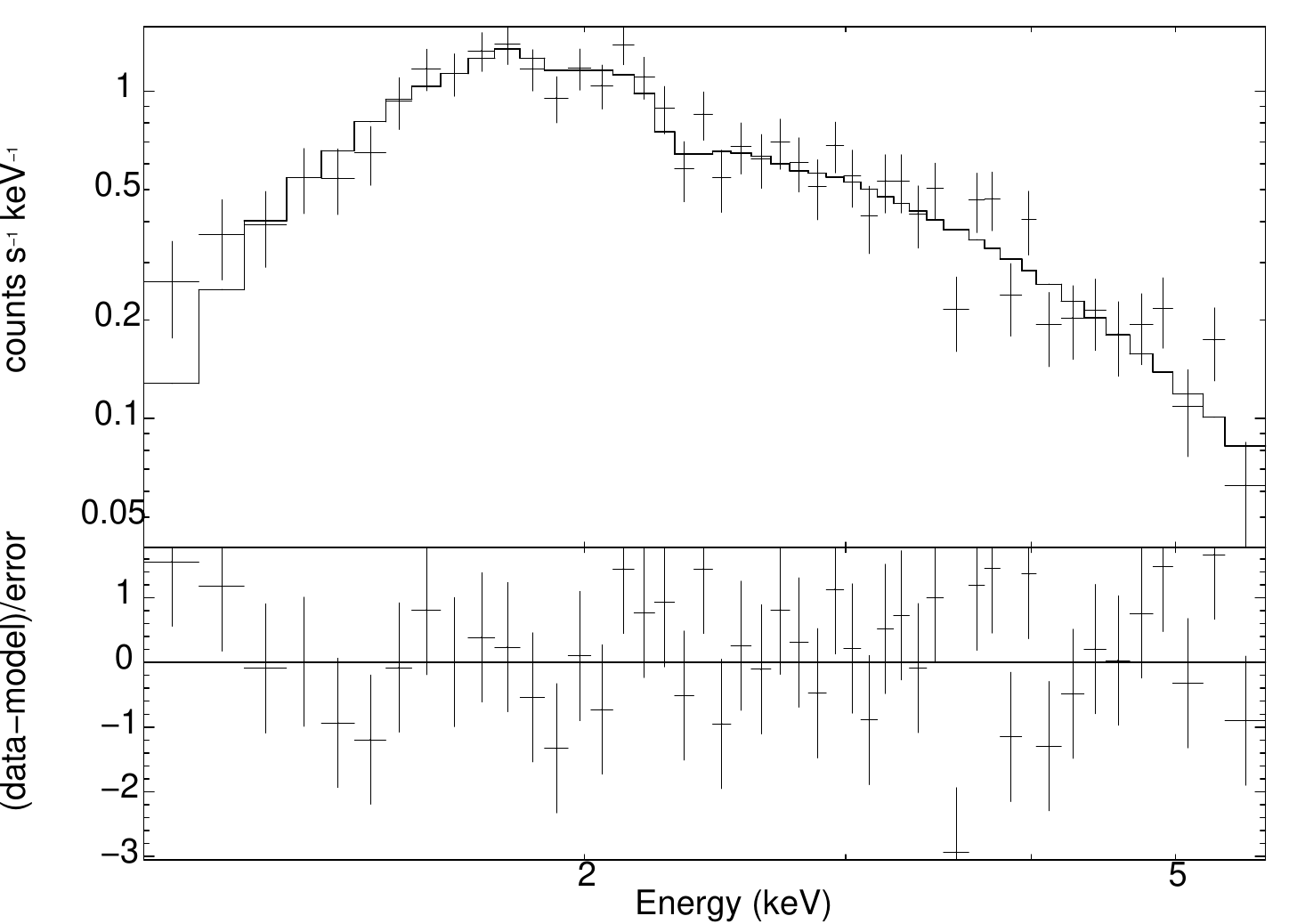}{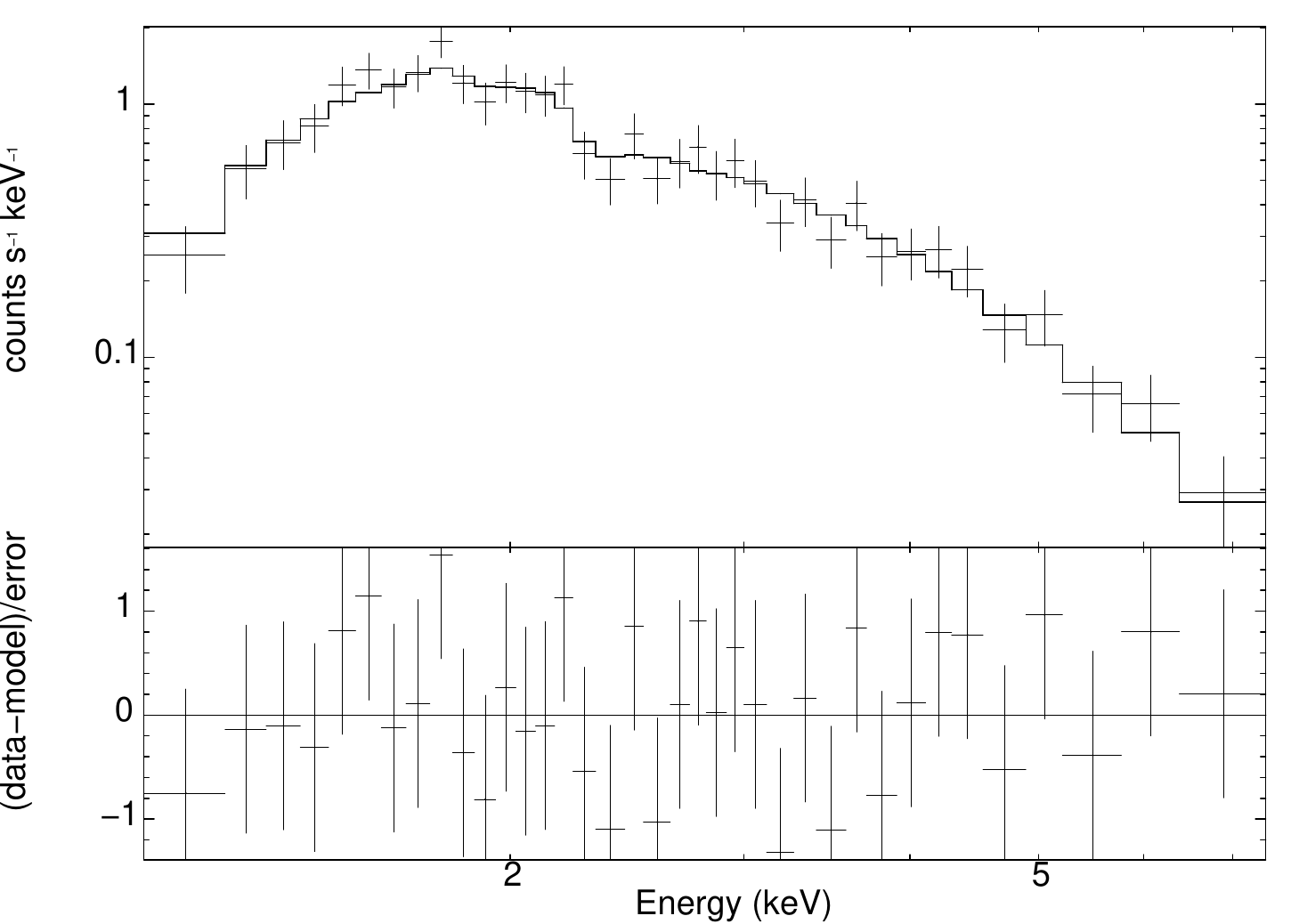}
        \caption{NICER spectra (during quiescent state) fitted with the absorbed Comptonized model for 2024 April 24 (MJD 60424.4) and 2024 April 25 (MJD 60425) are shown in the left and right panels, respectively. The residuals between the data and the model are shown in the lower panel in each plot.}
        \label{f4}
\end{figure*}

We carry out spectral fitting of the NICER spectra obtained during the quiescent state (Obs. IDs 6639080116, 6639080133 and 6639080134) on 2024 March 23 (MJD 60392.1), 2024 April 24 (MJD 60424.4) and 2024 April 25 (MJD 60425). The NICER spectra obtained for other epochs after 2024 March 23 (MJD 60392.1) are either nearly background dominated or have poor statistics for spectral fitting. The results of our spectral analysis during quiescent state are summarized in Table \ref{table:spec_2}. The spectra for MJD 60424.4 was fitted in the 1-6 keV energy band with the absorbed thermal Comptonized model (Figure \ref{f4}).
The fitted hydrogen column density was $N_{\rm H} = (3.1\pm0.5)\times 10^{22}$ cm$^{-2}$. We obtained photon-index $\Gamma$ = 2.1$\pm$0.4, with $\chi^{2}$ = 45.1, d.o.f. = 43, and reduced $\chi^{2}$ = 1.1 (Table \ref{table:spec_2}). The inferred electron temperature ($kT_e$) was $4.2\pm1.1$ keV. The temperature of the blackbody seed photons ($kT_{bb}$) was fixed at 0.1 keV.
The unabsorbed flux in the energy range 0.5-10 keV
 is $6.3 \pm 1.5 \times10^{-11}$ erg cm$^{-2}$ s$^{-1}$, which implies a source luminosity of $7.5 \pm 1.8 \times10^{35}$ erg s$^{-1}$ for a distance of $\sim$10 kpc. The estimated luminosity in the 0.5-100 keV energy range is $7.9 \pm 1.8 \times10^{35}$ erg s$^{-1}$. The spectra for this epoch could also be described using an absorbed power-law ({\tt tbabs$\times$(powerlaw)}) model. We obtained hydrogen column density $N_{\rm H} = (3.2\pm0.5)\times 10^{22}$ cm$^{-2}$ and a photon-index $\Gamma$ = 2.2$\pm$0.3,  with $\chi^{2}$ = 45.2, d.o.f. = 44, and reduced $\chi^{2}$ = 1.03 (Table \ref{table:spec_2}). The estimated absorption and photon-index are broadly consistent with those obtained using the absorbed thermal Comptonized model. The unabsorbed flux is $6.9 \pm 1.1 \times10^{-11}$ erg cm$^{-2}$ s$^{-1}$  in the energy range of 0.5-10 keV, which implies a source luminosity of $8.3 \pm 1.3 \times10^{35}$ erg s$^{-1}$ for a distance of $\sim$10 kpc. The spectra during the quiescent state, fitted using the absorbed thermal Comptonized model, shows a soft excess (Figure \ref{f4}), which can be reduced using the absorbed black body ({\tt tbabs$\times$(bbodyrad)}) model, yielding hydrogen column density $N_{\rm H} = (1.5\pm0.3)\times 10^{22}$ cm$^{-2}$ and a temperature $kT_{\rm bbody}$ of $0.85\pm0.07 $, with $\chi^{2}$ = 52.85, d.o.f. = 44, and reduced $\chi^{2}$ = 1.2 (Table \ref{table:spec_2}). The unabsorbed flux is $2.7\pm0.2 \times10^{-11}$ erg cm$^{-2}$ s$^{-1}$  in the energy range of 0.5-10 keV, which implies a source luminosity of $3.3\pm0.2 \times10^{35}$ erg s$^{-1}$ for a distance of $\sim$10 kpc.

  \begin{table*}[hbtp]
\begin{ruledtabular}
\centering
 \caption{\label{table:spec_2} Best-fit spectral parameters of the NICER data for SRGA J144459.2-60420 during the quiescent state.}
 \centering
 \begin{tabular}{llll}
Parameter (units) & {\tt tbabs$\times$(nthcomp)}& {\tt tbabs$\times$(powerlaw)}& {\tt tbabs$\times$(bbodyrad)}\\

  \hline
Epoch & MJD 60424.4   \\
 \hline

 % \hline
$N_{\rm H}~(10^{22}~ {\rm cm}^{-2})$& $ 3.1\pm0.5 $  & $ 3.2\pm0.5 $ &  $1.5\pm0.3 $\\
$\Gamma$                            & $ 2.1 \pm0.4 $      & $ 2.2 \pm0.3 $ & ...\\
$kT_{\rm e}$ (keV)                        &  $4.2\pm1.1$       & ... &...\\
$kT_{\rm bb}$ (keV)                       & 0.1 (fixed)      & ... & ...\\
  $kT_{\rm bbody}$ (keV)                & ... &     ...       & $ 0.85\pm0.07 $  \\
% \hline
 $\chi^{2}/{\rm d.o.f.}$ &   45.1/43 &   45.2/44 & 52.85/44\\
 $F$ ($10^{-11}$ erg s$^{-1}$ cm$^{-2}$)\textsuperscript{a} & $6.3\pm1.5$ & $6.9\pm1.1$ &  $2.7\pm0.2$\\
  \hline
Epoch & MJD 60425\\

   \hline
$N_{\rm H}~(10^{22}~ {\rm cm}^{-2})$& $  2.8\pm0.8$  & $  3.0\pm0.5$ & $ 1.3\pm0.3$\\
$\Gamma$                            & $ 2.1 \pm0.3 $      & $  2.3\pm0.1$ & ...\\
$kT_{\rm e}$ (keV)                        &  3 (fixed)       &  ...&  ...\\
$kT_{\rm bb}$ (keV)                       &   $  0.2\pm0.1$    &  ... &  ...\\
  $kT_{\rm bbody}$ (keV)                & ... &     ...       &  $ 0.86\pm0.08 $ \\
% \hline
 $\chi^{2}/{\rm d.o.f.}$ &   19.35/33 &   19.1/34 & 40.69/34\\
 $F$ ($10^{-11}$ erg s$^{-1}$ cm$^{-2}$)\textsuperscript{a} & $3.6\pm1.7$ & $4.7\pm1.4$ &  $1.8\pm0.2$\\
 % \hline

\end{tabular}
\end{ruledtabular}
\begin{tablenotes}
\item[a]  \textsuperscript{a} Unabsorbed flux in the 0.5--10 keV energy range.
\end{tablenotes}
 \label{tab:broadband_spec}
\end{table*}

 We tried fitting the spectra for 2024 April 25 (MJD 60425) using the absorbed thermal Comptonized model (Figure \ref{f4}). The spectral fitting was restricted to the energy band of 1-8 keV as the spectrum was background-dominated beyond this energy range.
 The fitted hydrogen column density was $N_{\rm H} = (2.8\pm0.8)\times 10^{22}$ cm$^{-2}$ and the photon-index $\Gamma = 2.1\pm0.3$, with $\chi^{2}$ = 19.35, d.o.f. = 33, and reduced $\chi^{2}$ = 0.6 (Table \ref{table:spec_2}). The unabsorbed flux in the energy range 0.5-10 keV
 is $3.6 \pm 1.7 \times10^{-11}$ erg cm$^{-2}$ s$^{-1}$, which implies a source luminosity of $4.3 \pm 2.0 \times10^{35}$ erg s$^{-1}$ for a distance of $\sim$10 kpc. The estimated luminosity in the 0.5-100 keV energy range is $4.7 \pm 1.8 \times10^{35}$ erg s$^{-1}$. The spectra were also fitted using an absorbed power-law model. We obtained hydrogen column density $N_{\rm H} = (3.0\pm0.5)\times 10^{22}$ cm$^{-2}$ and a photon-index $\Gamma = 2.3\pm0.1$, with $\chi^{2}$ = 19.1, d.o.f. = 34, and reduced $\chi^{2}$ = 0.6 (Table \ref{table:spec_2}).
 The unabsorbed flux is $4.7 \pm 1.4 \times10^{-11}$ erg cm$^{-2}$ s$^{-1}$  in the energy range of 0.5-10 keV, which implies a source luminosity of $5.7 \pm 1.7 \times10^{35}$ erg s$^{-1}$ for a distance of $\sim$10 kpc. The estimated luminosity in the 0.5-100 keV energy range is $\sim 6.4 \times10^{35}$ erg s$^{-1}$. The spectra during the quiescent state was also fitted using the absorbed black body model, yielding hydrogen column density $N_{\rm H} = (1.3\pm0.3)\times 10^{22}$ cm$^{-2}$ and a temperature $kT_{\rm bbody}$ of $0.86\pm0.08 $, with $\chi^{2}$ = 40.69, d.o.f. = 34, and reduced $\chi^{2}$ = 1.2 (Table \ref{table:spec_2}). The unabsorbed flux is $1.8\pm0.2 \times10^{-11}$ erg cm$^{-2}$ s$^{-1}$  in the energy range of 0.5-10 keV, which implies a source luminosity of $2.2\pm0.2 \times10^{35}$ erg s$^{-1}$ for a distance of $\sim$10 kpc.

\begin{figure*}
\centering
\plotone{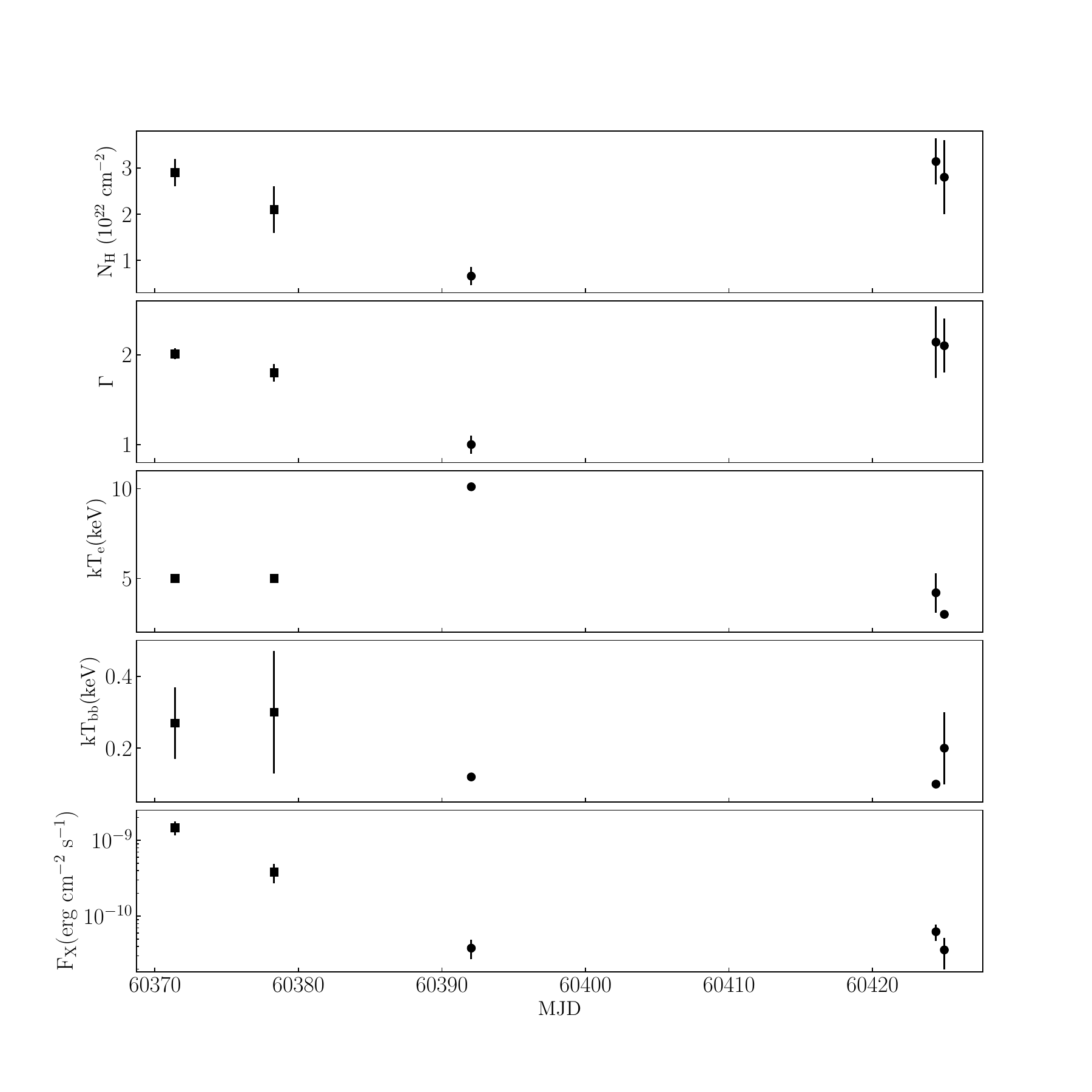}
  \caption{Plot showing the best-fitted parameters of the Swift (filled squares) and NICER spectra (filled circles) using the absorbed Comptonized model. This figure is only showing outburst decay and quiescence, with the reflare observations omitted.
  The hydrogen column density, $\Gamma$, the electron
temperature, the blackbody temperature and the unabsorbed flux in the 0.5-10 keV energy band are shown in the panels from top to bottom.}
 \label{f5}
\end{figure*}

\subsection{Spectral evolution}
The spectral evolution of SRGA J144459.2-60420 during the outburst decay and quiescent state is shown in Figure \ref{f5}, which includes results from both the Swift/XRT and NICER fits.
It is observed that, in general, the absorption is higher than the interstellar absorption ($N_{\rm H} \sim 1.7 \times 10^{22}$ cm$^{-2}$) along the direction of this source \citep{bekhti2016hi4pi}. The spectral index ($\Gamma$) has no indication of monotonic variation during the outburst decay and the quiescent state. There is an indication of decrease in the electron temperature of the corona during the quiescent phase of the source. The temperature of the blackbody seed photons is $\sim$0.3 keV and gradually decreases to $\sim$0.1 keV during the quiescent state (around MJD 60424.4). The inferred unabsorbed flux in the energy range of 0.5-10 keV decreases from $\sim 1.5 \times10^{-9}$ erg cm$^{-2}$ s$^{-1}$ (MJD 60371.4) to $\sim 3.8 \times10^{-11}$ erg cm$^{-2}$ s$^{-1}$ (MJD 60392.1). The flux drops by $\sim$40 times during this period. The flux drops by $\sim$4 times during the period between MJD 60371.4 and 60378.3.
Thereafter, the decay of the outburst persists as the flux drops by $\sim$10 times during the period between MJD 60378.3 and 60392.1.
The flux inferred approximately after a month (MJD 60424.4) is $\sim 6.3 \times10^{-11}$ erg cm$^{-2}$ s$^{-1}$, which is higher than that obtained around
MJD 60392.1 by a factor of $\sim$1.7.
Thereafter, the flux decreases to $\sim 4.7 \times10^{-11}$ erg cm$^{-2}$ s$^{-1}$ within a day.

\subsection{Long-term quiescent X-ray history of SRGA J144459.2-60420}
SRGA J144459.2-60420 was in the field of view of the Einstein Observatory on 1979 August 26 (MJD 44111.3) for $\sim$2.8 ks. The estimated upper limit flux in the 0.2-2 keV energy band is $2.9 \pm 1.3 \times10^{-13}$ erg cm$^{-2}$ s$^{-1}$ obtained using the High-Energy Lightcurve Generator (HILIGT) online tool\footnote{\url{https://xmmuls.esac.esa.int/upperlimitserver/}} \citep{konig2022hiligt}, assuming a $\Gamma$ = 2 power law to convert from count rate to flux in the 0.2-2 keV energy band. The upper limits obtained from other observatories are also obtained using the HILIGT tool assuming a $\Gamma$ = 2 power law. The detailed process of computing upper limits is given in  \citet{konig2022hiligt}. The estimated upper limit flux of the source in the 0.2-2 keV energy band from the ROSAT survey (1990 July 30, MJD 48102) for 0.4 ks exposure was $\sim 3.9 \times10^{-13}$ erg cm$^{-2}$ s$^{-1}$. The source was observed during the ROSAT/PSPC pointed observations on 1992 September 2 (MJD 48867.2) and 1993 September 14 (MJD 49244.9) for $\sim$2 ks and 0.2 ks, respectively. The estimated upper flux limits in the 0.2-2 keV energy band were $\sim 2.2 \times10^{-13}$ erg cm$^{-2}$ s$^{-1}$ and $\sim 1.1 \times10^{-12}$ erg cm$^{-2}$ s$^{-1}$, respectively.  The source lay in the field of view of the INTEGRAL observatory for $\sim$1100 ks on 2003 January 29 (MJD 52668.4). The estimated hard X-ray flux in the 20-40 keV energy band was $\sim 1.1 \times10^{-12}$ erg cm$^{-2}$ s$^{-1}$. The source was observed on two occasions on 2010 December 16, MJD 55546.2 and 55546.9 by the Swift/XRT having exposures of $\sim$0.4 ks and 0.6 ks, respectively. The upper limit flux in the 0.2-12 keV energy range obtained during these two observations, separated by $\sim$16 hr, were  $\sim 1.1 \times10^{-12}$ erg cm$^{-2}$ s$^{-1}$ and $\sim 5.9 \times10^{-13}$ erg cm$^{-2}$ s$^{-1}$, respectively. The source happened to lie in the XMM-Newton slew observations on 2011 August 13 (MJD 55786.3) and 2013 March 3 (MJD 56354.1) for $\sim$6 s and 2 s, respectively. The inferred upper limits in the 0.2-12 keV energy band were $\sim 3.8 \times10^{-12}$ erg cm$^{-2}$ s$^{-1}$ and $\sim 1.3 \times10^{-11}$ erg cm$^{-2}$ s$^{-1}$ respectively. The estimated upper limit flux from eROSITA 0.1 ks observations on 2020 February 29 (MJD 58908) was $\sim 1.8 \times10^{-12}$ erg cm$^{-2}$ s$^{-1}$. The estimated upper-limit fluxes from various observatories are shown in Table \ref{tab3}.

The typical upper limit flux of the source in the soft X-ray energy band ($\sim$0.2-10 keV) spans the range of $\sim 3 \times10^{-13}$ erg cm$^{-2}$ s$^{-1}$ to $\sim 1.8 \times10^{-12}$ erg cm$^{-2}$ s$^{-1}$. This translates to the source luminosity range of $\sim 0.4-2.2 \times10^{34}$ erg s$^{-1}$ for a distance of $\sim$10 kpc. The coronal activity of the companion star alone cannot power this luminosity, as the typical X-ray luminosity from late-type low-mass stars is $\sim 10^{32}$ erg s$^{-1}$
\citep{dempsey1993rosat,verbunt1996rosat}.

\begin{table*}
\centering
\caption{Estimated upper-limit fluxes for SRGA J144459.2-60420 from various archival observations.
\label{tab3}}
\begin{tabular}{lccc}
\hline\hline
MJD & Observatory & Energy range (keV) & Flux ($10^{-13}$ erg cm$^{-2}$ s$^{-1}$)\\
            \hline
44111.3	&	Einstein	&	0.2-2	&	$\sim$2.9	\\
48102	&	ROSAT	&	0.2-2	&	$\sim$3.9	\\
48867.2	&	ROSAT	&	0.2-2	&	$\sim$2.2	\\
49244.9	&	ROSAT	&	0.2-2	&	$\sim$11	\\
52668.4	&	INTEGRAL	&	20-40	&	$\sim$11	\\
55546.2	&	Swift	&	0.2-12	&	$\sim$11	\\
55546.9	&	Swift	&	0.2-12	&	$\sim$5.9	\\
55786.3	&	XMM-Newton	&	0.2-12	&	$\sim$3.8	\\
56354.1	&	XMM-Newton	&	0.2-12	&	$\sim$130	\\
58908	&	eROSITA	&	0.2-12	&	$\sim$18	\\
\hline

\end{tabular}
\end{table*}

We compute the quiescent thermal luminosity of the source using the deep crustal heating model \citep{brown1998crustal} to probe if the long-term quiescent upper limits obtained during the period spanning 1979-2023 can be explained using this model. In this model, the neutron star is heated due to a chain of nuclear reactions that occur in the crust during outbursts. The neutron star's soft X-ray emission during quiescence may be powered by the heat radiated from the crust. The quiescent thermal luminosity in this model is given by \citep{brown1998crustal,degenaar2012quiescent},

\begin{equation}
 L_{\mathrm{th,bol}} = \langle \dot{M} \rangle ~\frac{Q_{\mathrm{nuc}}}{m_{\mathrm{u}}} \simeq 1.9 \times 10^{18} ~\langle \dot{M} \rangle,
\end{equation}
where $m_{\mathrm{u}}$ is the atomic mass unit, $Q_{\mathrm{nuc}}\simeq2$~MeV is the energy released in the crust per accreted nucleon \citep{gupta2007heating,haensel2008models}, and $\langle \dot{M} \rangle$ is the mass accretion rate averaged over $\simeq$$10^4$~years. $\langle \dot{M} \rangle$ can be estimated using $\langle \dot{M} \rangle \sim \langle \dot{M}_{\mathrm{obs}} \rangle \times \frac{t_{\mathrm{out}}}{t_{\mathrm{recur}}}$ \citep{degenaar2012quiescent}, where $\langle \dot{M}_{\mathrm{obs}} \rangle $ is the average
accretion rate during outburst, $t_{\mathrm{out}}$ is the average outburst duration, and $t_{\mathrm{recur}}$ is the recurrence time. SRGA J144459.2-60420 was detected in a faint state in the 2-10 keV energy band from archival MAXI observations in 2022 January and 2023 December \citep{negoro2024maxi}. The source was also detected in 2023 December in the 25-60 keV energy range using archival INTEGRAL observations \citep{sguera2024integral}. The 2024 outburst lasted for approximately 30 days. Therefore, the estimated $t_{\mathrm{recur}}$ is $\sim$1.04 years. The mean MAXI one-day averaged count rate during the outburst duration \citep[MJD 60355-60385][]{li2025timing} was $\sim$0.18 $\mathrm{Photons ~cm^{-2} ~s^{-1}}$, while during the outburst peak, it was $\sim$0.43 $\mathrm{Photons ~cm^{-2} ~s^{-1}}$. The peak bolometric flux (1-250 keV) during the outburst was $\sim 4 \times10^{-9}$ erg cm$^{-2}$ s$^{-1}$ \citep{li2025timing}. Scaling the peak flux by the average MAXI count rate during the outburst, assuming that the spectral shape remains the same in every epoch, the averaged outburst flux ($\langle \dot{M}_{\mathrm{obs}} \rangle $) is estimated to $\sim 1.6 \times10^{-9}$ erg cm$^{-2}$ s$^{-1}$. Using $t_{\mathrm{out}} \sim$30 days and $t_{\mathrm{recur}}\sim$1.04 year, we obtain $\langle \dot{M} \rangle \sim 8.4 \times10^{15}$ g s$^{-1}$. The quiescent thermal luminosity is estimated to  $L_{\mathrm{th,bol}} \sim 1.6 \times10^{34}$ erg s$^{-1}$. The estimated quiescent thermal luminosity using the deep crustal heating model is comparable to the minimum upper limit on the luminosity estimated in the 0.2-10 keV energy band during the quiescent state of the source ($\sim 0.4-2.2 \times10^{34}$ erg s$^{-1}$) during the period 1979-2023.
It should be noted that the estimated average
accretion rate during outburst over timescales of $\simeq 10^4$~years may be different from that estimated during the 2024 outburst. The mass-transfer rate averaged over the lifetime of an accreting binary ($\simeq 10^9$~years) consisting of a Roche-lobe filling main-sequence star having an orbital period of $\sim$5 hr \citep{ng2024nicer} and donor mass lying in the range of 0.3-0.4 $M_{\odot}$ \citep{dohi2025evidence} is $\langle \dot{M} \rangle \sim 1.6 \times 10^{-11} ~M_{\odot} \mathrm{yr}^{-1}$ \citep{verbunt1993origin}.
This estimated mass accretion rate is roughly 8 times smaller than that estimated during the 2024 outburst of the source. It should be noted that the mass accretion rate averaged over $\simeq$$10^4$~years computed in this paper depends on $t_{\mathrm{out}}$ and $t_{\mathrm{recur}}$, which are not very well constrained in this source. On the other hand, mass-transfer rate averaged over the lifetime of an accreting binary ($\simeq 10^9$~years) assumes that the entire mass lost by the donor star is captured by the companion and there is no mass loss from the binary system. Therefore, the mass-transfer rates estimated using these methods are likely to be overestimated.
Thus, the long-term quiescent X-ray luminosity of the source may be powered using the quiescent thermal luminosity of the source generated using the deep crustal heating model. Future observations of the source during outbursts would enable us to further constrain the recurrence time of the outbursts in this source and verify the estimated limits on the quiescent thermal luminosity.

\section{Discussions}
 We find that the MAXI and the NICER light curve of the source gradually decays after the outburst, but shows a few reflares. The spectra during the outburst decay, reflares, and the quiescent state is explored using the absorbed Comptonized model. The estimated luminosity (0.5-10 keV energy band) during reflares ($\sim 1.6\times10^{36}$ erg s$^{-1}$) is an order of magnitude smaller than that obtained during the outburst decay ($\sim 1.8 \times10^{37}$ erg s$^{-1}$). In addition, the spectra during the quiescent state is also explored using the absorbed power-law and absorbed blackbody model having spectral index $\Gamma \sim$2.2-2.3 and temperature $kT_{\rm bbody} \sim$0.9 keV, respectively. The estimated luminosity during the quiescent state ($\sim 3.3-7.5 \times10^{35}$ erg s$^{-1}$) is approximately two to five times smaller than that obtained during reflares. The spectral evolution of the source during the outburst decay, reflares, and the quiescent state is investigated. We also explore the long-term quiescent X-ray history of SRGA J144459.2-60420 using archival observations and our findings indicate that the long-term quiescent X-ray luminosity in this source may be explained using the deep crustal heating model.

\subsection{X-Ray emission during the 2024 quiescent state}
The X-ray spectrum of SRGA J144459.2-60420 during the quiescent state was fitted using the absorbed thermal Comptonized model, absorbed power-law, and absorbed black body model. The absorption and the spectral index in the {\tt nthcomp} model was $\sim 3\times 10^{22}$ cm$^{-2}$ and $\sim 2.1$, respectively. The seed photon and electron temperature in the {\tt nthcomp} model during the quiescent state was similar to the seed photon temperature ($\sim$0.1-0.4 keV) and electron temperature ($\sim$2-7 keV) obtained during the outburst of the source \citep{li2025timing}. The absorption and the spectral index in the absorbed power law model was $\sim 3\times 10^{22}$ cm$^{-2}$ and $\sim 2.2-2.3$, respectively.  The temperature in the blackbody model during the quiescent state was $\sim$0.85 keV, which is comparable to $\sim$0.85-1 keV obtained during the outburst of the source \citep{ng2024nicer}. The blackbody radius estimated from the blackbody model during the quiescent state was $\sim$2 km, which is compatible with the presence of a hot spot on the neutron star surface.
The X-ray spectrum of SRGA J144459.2-60420 during the quiescent state becomes softer relative to the spectrum during the bright state of the outburst around 2024 February \citep{li2025timing}. The photon-index from spectral fitting is $\sim$2.1-2.3 during the quiescent state (2024 April 24-25, Table \ref{table:spec_2}) compared to that $\sim$2 or lower approximately a month earlier in 2024 March. A similar photon index of $\sim$2 was obtained during the bright state of the outburst around 2024 February \citep{li2025timing}.
The spectral softening near the end of the outburst/quiescence (Figure \ref{f5}) may be similar to that observed in both black hole \citep{tomsick2004detection,kalemci2005multiwavelength,corbel2006origin,wu2008x,dunn2010global,armas2013multiwavelength,plotkin2013x} and neutron star binaries \citep{campana1998neutron,armas2013x,chandra2025long}. It is plausible that residual
accretion onto the neutron star surface could explain the spectral softening near the end of the outburst/quiescence \citep{campana1998neutron,degenaar2012quiescent}. The quiescent spectra of accreting neutron stars have been described using the absorbed blackbody and power-law model \citep{degenaar2012quiescent,tsygankov2017x}. In addition, the quiescent spectra of AMXPs have also been fitted using the absorbed power-law model having $\Gamma \simeq$1-2 \citep{degenaar2012quiescent}, which is similar to our findings in SRGA J144459.2-604207 having $\Gamma \sim$2 (Table \ref{table:spec_2}). Other possible non-accretion mechanisms which can explain the non-thermal component of the power-law emission is the shock formed due to the collision between the pulsar wind and the matter emanating from the donor star, or due to the mechanism which produces the pulsar wind \citep{campana1998neutron}. The relativistic pulsar wind composed of electrons and positrons may collide with nearby gas or leftover matter from the accretion disc and form a shock. This shock can cause particle acceleration and lead to synchrotron emission, which may manifest as a power law in the X-ray band.
The coronal activity of the companion star can contribute to the quiescent X-ray luminosity, but alone cannot power X-ray luminosity of $\sim 10^{35}$ erg s$^{-1}$ as late-type low-mass stars have X-ray luminosity of $\sim 10^{32}$ erg s$^{-1}$
\citep{dempsey1993rosat,verbunt1996rosat}.

\begin{figure*}
\centering
\plotone{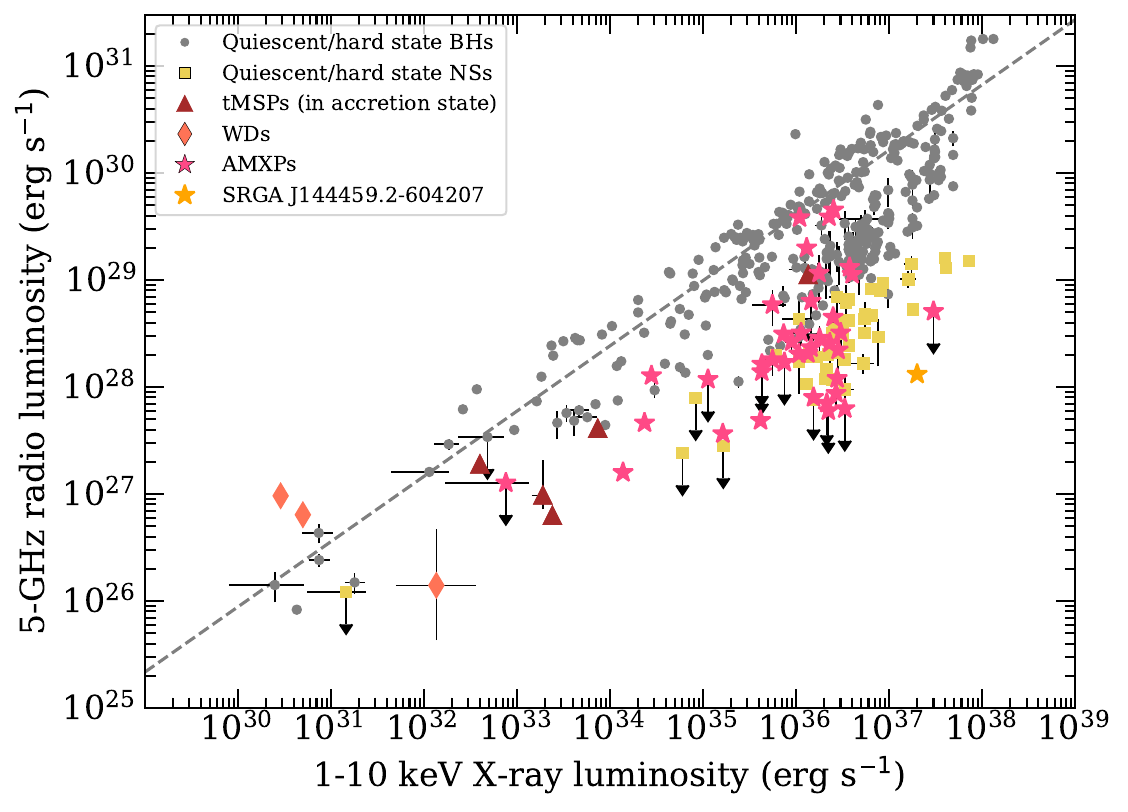}
  \caption{X-ray (1-10 keV) luminosity versus radio (5 GHz) luminosity for  LMXBs (containing black hole (BH) and neutron star (NS)), transitional millisecond pulsars (tMSPs), white dwarfs (WDs), and AMXPs. SRGA J144459.2-604207 is marked by an orange star in the Radio-X-ray  luminosity plane.
  The gray circles represent
BHs from the literature \citep{gallo2003universal,corbel2003radio,merloni2003fundamental,gallo2006radio,miller2011deep,gallo2012assessing,strader2012two,chomiuk2013radio,gallo2014radio,ratti2012black,corbel2013universal,miller2015deep,rushton2016disc,ribo2017first,plotkin2017up,dinccer2017multiwavength,motta2018radio,carotenuto2021hybrid,carotenuto2022black}. LMXBs \citep{tetarenko2016disc,gusinskaia2017jet,tetarenko2018radio}, AMXPs \citep{miller2010radio,migliari2011influence,tudor2016radio,tudor2017disc,tetarenko2018radio,russell2018radio,van2018vla,gusinskaia2020radio}, WDs \citep{marsh2016radio}, and tMSPs \citep{hill2011bright,papitto2013swings,deller2015radio,bogdanov2018simultaneous}
  from the literature are marked by squares, stars, diamonds, and triangles, respectively. The BH population slope ($\beta = $ 0.61, \citet{gallo2014radio}) is marked with a dashed line.}
 \label{f6}
\end{figure*}

\subsection{Radio–X-Ray relation}
We explore the radio vs X-ray relation of the source by placing it on the radio-X-ray plane using the 5 GHz radio luminosity and the 1.0-10 keV X-ray luminosity estimated during the hard spectral state of the X-ray outburst of the source. The source was detected using the Australia Telescope Compact Array (ATCA) observations on 2024 February 29 (MJD 60369.6) at 5.5 GHz and 9 GHz \citep{russell2024atca}. The measured flux density at 5.5 GHz was $200 \pm 15 ~\mu$Jy \citep{russell2024atca}, which implies radio luminosity of $1.3 \pm 0.1 \times10^{28}$ erg s$^{-1}$, assuming a distance of $\sim$10 kpc. The estimated X-ray flux (1-250 keV energy band) during the period MJD 60367.0-60369.88 (2024 February 27-29) was $\sim 3.8 \times10^{-9}$ erg cm$^{-2}$ s$^{-1}$ \citep{li2025timing}. We estimated the X-ray flux in the 1-10 keV energy band using the WebPIMMS\footnote{\url{https://heasarc.gsfc.nasa.gov/cgi-bin/Tools/w3pimms/w3pimms.pl}} tool to $\sim 1.8 \times10^{-9}$ erg cm$^{-2}$ s$^{-1}$, assuming an absorbed power-law model with $\Gamma \sim$2 and interstellar absorption ($N_{\rm H} \sim 1.7 \times 10^{22}$ cm$^{-2}$). This X-ray flux conversion using WebPIMMS assumes that there is no spectral break in the broadband spectrum of the source. The luminosity estimated in the 1-10 keV energy band is $\sim 2  \times10^{37}$ erg s$^{-1}$ assuming distance of $\sim$10 kpc. The location of SRGA J144459.2-60420 in the radio–X-ray plane is shown in Figure \ref{f6} along with LMXBs (containing black hole (BH) and neutron
star), transitional millisecond pulsars (tMSPs), white dwarfs (WDs), and other AMXPs.
For BH-LMXBs, there exists a power law relation between the radio ($L_R$) and X-ray ($L_X$) luminosities in the hard X-ray state ($L_R \propto {L}_X^\beta$), which is shown by a dashed line in Figure \ref{f6} for $\beta = $ 0.61 \citep{gallo2014radio}. The location of SRGA J144459.2-604207 in the radio-X-ray plane lies in the region of parameter space occupied by AMXPs, which broadly span the range of X-ray and 5 GHz radio luminosity of $\sim 10^{34-38}$\,erg\,s$^{-1}$ and  $\sim 10^{27-29}$\,erg\,s$^{-1}$, respectively. It is seen from Figure \ref{f6} that both non-pulsating neutron star LMXBs and AMXPs
can show a range of radio luminosities for similar X-ray luminosity. Similarly, for a given radio luminosity of $\sim 10^{28}$\,erg\,s$^{-1}$ for AMXPs, the X-ray luminosities vary by roughly three orders of magnitude with SRGA J144459.2-604207 lying on the higher end.

\subsection{Reflares during outburst decay}
\begin{figure}
\centering
\plotone{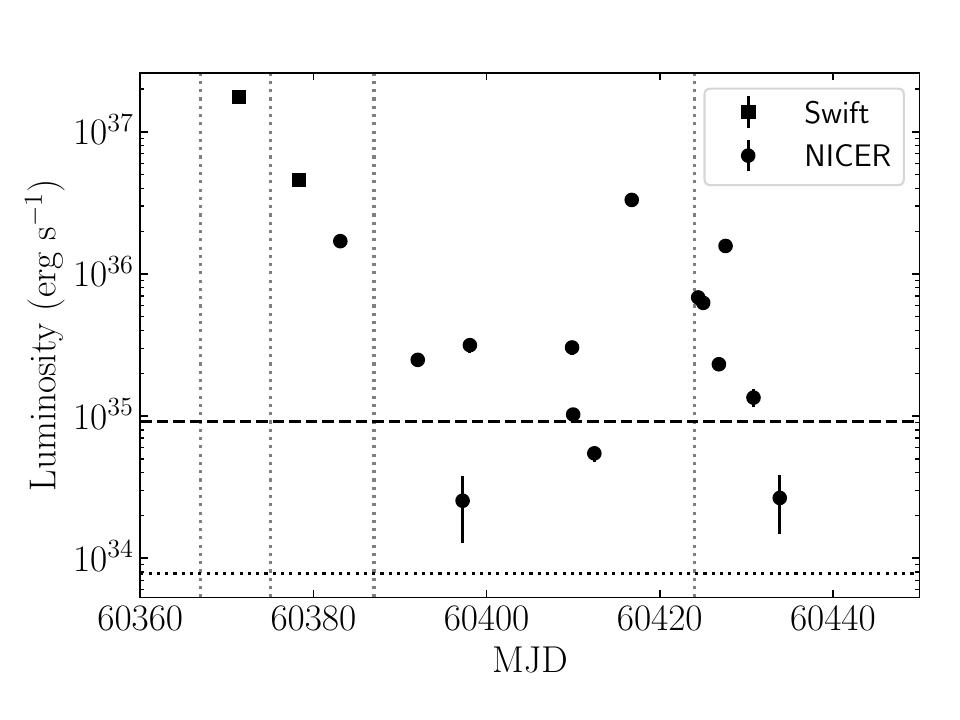}
  \caption{Plot showing estimated X-ray luminosity (0.5-100 keV) of SRGA J144459.2-604207 spanning the duration 2024 March 2 (MJD 60371.4) until 2024 May 3 (MJD 60433.9). The horizontal dashed and dotted lines show the estimated limiting luminosity for the propeller effect to set in for B=$1.3 \times 10^{8}$ G and B=$3.8 \times 10^{7}$ G, respectively. The vertical dotted lines show the epochs when reflares were observed in the MAXI one-day averaged light curve.}
 \label{f7}
\end{figure}

Reflares were observed around MJD 60367, MJD 60375, MJD 60387, and MJD 60424, lasting for at most a few days in the MAXI one-day averaged light curve (2-20 keV energy band) of SRGA J144459.2-604207 (Figure \ref{f1}). The reflare around MJD 60367 was also detected using NICER and Insight-HXMT observations \citep{li2025timing}. Reflares have been detected in accreting NS binaries such as SAX J1808.4-3658 \citep{patruno2016reflares}, MAXI J1807+132 \citep{jimenez2019complex}, and Swift J1858.6-0814 \citep{buisson2021dips}. It should be noted that the flaring behaviour in Swift J1858.6-0814 was due to variable obscuration \citep{buisson2021dips} and the flaring behaviour observed in SRGA J144459.2-604207 is not likely to be caused by a similar phenomenon. The typical flare duration in SAX J1808.4-3658 and MAXI J1807+132 is $\sim$1-2 days and $\sim$2 days \citep{rout2025multi,patruno2016reflares}, respectively, which is similar to that observed in SRGA J144459.2-604207.
In MAXI J1807+132, the decay of the 2017 outburst was characterized by the presence of several reflares having quasi-periodic recurrence time of $\sim$6.5 days \citep{jimenez2019complex}.
We now compare the estimated luminosity before and after the reflares with the propeller luminosity.
The limiting luminosity for the onset of the propeller effect ($L_{\rm prop}$) is given by \citep{campana1998aquila,tsygankov2016propeller,chandra2025long},

\begin{equation}
  L_{\rm prop}
\simeq 4 \times 10^{37} ~k^{7/2}
~B_{12}^2 ~P_s^{-7/3} ~M_{1.4}^{-2/3} ~R_6^5 \,\textrm{erg s$^{-1}$} ,\label{eq2}
\end{equation}
where M, R, $P_s$ and B are the mass, radius, spin period and magnetic field of the neutron star, respectively.
The factor $k$ is the ratio of the magnetospheric radius and the Alfv\'en radius, which in the case of disc accretion is taken to be $k=0.5$ \citep{ghosh1978disk}. The upper limit on the magnetic field can be estimated by equating the magnetospheric
(Alfv\'{e}n) radius ($r_{\rm m}$) and the co-rotation radius of the neutron star ($r_{\rm co}$). The Alfv\'{e}n radius is given by \citep{chandra2026},

\begin{equation}
r_{\rm m}=1.6\times 10^{8} ~\xi ~\dot{M}^{-2/7}_{18}M^{-1/7}_{1.4}\mu^{4/7}_{30}~\rm cm,
\end{equation}
where $\dot{M}_{18}$, $M_{1.4}$ and $\mu_{30}$ are in units of $10^{18}~\rm g\,s^{-1}$, $1.4~\rm M_{\odot}$, and $10^{30}~\rm G\,cm^{3}$ respectively. We have used $M=1.4 ~M_{\odot}$ and R = 10 km throughout the paper. The co-rotation radius is given by \citep{chandra2026},

\begin{equation}
r_{\rm co} =\biggl( {\frac{GMP_s^{2}}{4\pi^{2}}} \biggr)^{1/3}~\rm cm. \label{eq4}
\end{equation}
Using $P_s \sim$2.23 ms \citep{ng2024nicer}, the estimated co-rotation radius is $\sim$28.7 km. Assuming $r_{\rm m}=r_{\rm co}$, the upper limit on the magnetic field ($B_{\mathrm{s,max}}$) is obtained as,

\begin{equation}
    B_{\mathrm{s,max}} < \xi^{-7/4} ~r_{\mathrm{co}}^{7/4} ~(2\,G\,M\,\dot{M}^2)^{1/4} ~R^{-3} \, \mathrm{G}.
\end{equation}
Using $\xi=0.5$ (for disc accretion, \citet{ghosh1978disk}), $r_{\rm co} \sim$28.7 km, and $\dot{M} \sim 3.26 \times 10^{-9} ~M_{\odot} ~\mathrm{yr}^{-1}$, we obtain $B_{\mathrm{s,max}} \sim 1.3\times 10^8$ G. The \textit{NICER} count rates shown in Figure {\ref{f2}} are converted into the corresponding 0.5-100 keV luminosities (shown in Figure {\ref{f7}}) by scaling the observed count rates using the luminosity ($\sim 1.7\times 10^{36}$\,erg\,s$^{-1}$, 0.5-100 keV energy band) for 2024 March 14 (MJD 60383.1) obtained earlier. An assumption is made during this conversion that the spectrum of the pulsar does not change significantly. Using $P_s \sim 2.23$ ms \citep{ng2024nicer}, B = $1.3 \times 10^{8}$ G and k = 0.5 (assuming disc accretion) in equation \ref{eq2}, the estimated $L_{\rm prop}$ is $\sim 9.1\times 10^{34}$\,erg\,s$^{-1}$, which is shown by a dashed horizontal line in Figure \ref{f7}.  The lower limit on the magnetic field obtained by \citet{li2025timing} is B = $3.8 \times 10^{7}$ G, which gives $L_{\rm prop} \sim 7.8\times 10^{33}$\,erg\,s$^{-1}$ and is shown by a dotted horizontal line in Figure \ref{f7}.
The inferred luminosity before and after the reflares around MJD 60367, and MJD 60375 is $\sim 10^{36-37}$ erg s$^{-1}$ (Figure \ref{f7}), which is larger than the estimated luminosity for the propeller effect to set in.

\subsection{Presence of jet/outflow during reflare?}
A blueshifted absorption feature $\sim$9.7 keV (interpreted as Fe XXVI edge) was detected in the XMM-Newton observations around 2024 February 28-29 (MJD 60368.1-60369.5), which was interpreted to be due to an ultrafast ($\simeq$ 0.04 c, $\rm{v}\sim$13000 $\rm{km ~s^{-1}}$) outflow emanating from the binary \citep{malacaria2025disk}. The detection of the outflow was nearly simultaneous with the peak of a reflare detected around 2024 February 27 (MJD 60367) and this reflare began around 2024 February 25 (MJD 60365) \citep{li2025timing}. In addition, radio emission at 5.5 GHz and 9 GHz was detected from the object using the Australia Telescope Compact Array (ATCA) observations on 2024 February 29 (MJD 60369.45-60369.82) \citep{russell2024atca}, which was nearly simultaneous with the detection of an ultrafast outflow from the object. The estimated radio spectral index of $\sim$-0.3 was suggested to be due to a compact
jet or discrete ejecta from the accreting pulsar \citep{russell2024atca}. This section corroborates this claim by using the X-ray data and making some calculations.
The detection of pulsations from the source using NICER, NuStar, IXPE, and XMM-Newton observations during this period \citep{papitto2025discovery} suggests that accreted material reaches the surface of the neutron star. The condition for jet formation requires that the gas pressure dominates the magnetic pressure and is given by \citep{massi2008magnetic},

\begin{equation}
\frac{r_{\rm m}}{R_*} \simeq  0.87 \left ( { B_*\over 10^8 ~ \rm G} \right)^{4/7} \left ( {\dot M\over {10^{-8}~ {\rm{M}_{\odot} ~yr^{-1}}}} \right)^{-2/7},
\label{eq6}
\end{equation}
where $r_{\rm m}$ is the Alfv\'en radius, $R_*$, and $B_*$ are the radius and magnetic field of the neutron star, respectively. $\frac{r_{\rm m}}{R_*} \approx 1$ covers the parameter space where the jet formation is not suppressed by the magnetic field \citep{massi2008magnetic,tetarenko2018radio}. In case of neutron stars LMXBs, the inner accretion disc may be truncated at the magnetospheric radius, where the ram pressure due to the infalling matter is balanced due to the magnetic pressure \citep{cackett2009broad}. Assuming that the accretion disc is truncated at the magnetospheric radius, the magnetic dipole moment of the neutron star ($\mu$) can be estimated using \citep{ibragimov2009accreting,cackett2009broad},

\begin{equation}
 \mu =\; 3.5 \times 10^{23} \, x^{7/4} \, k_A^{-7/4}\,M_{1.4}^2 \nonumber \\
         \times \left( \frac{f_{\rm ang}}{\eta} {F}_{b,*} \right)^{1/2}
\left( \frac{D}{3.5\, \mathrm{kpc}} \right),
\end{equation}
where $k_A$ is the geometric coefficient, $\rm f_{ang}$ is anisotropy correction factor accounting for the angular anisotropy of radiation and general relativity effects  \citep{ibragimov2009accreting}, and $\eta$ is the accretion efficiency.
$x$ is the scaling factor, which can be estimated from $\text R_{\rm in} = \frac{xGM}{c^2}$, where $\text R_{\rm in}$ is the inner disc radius. ${F}_{b,*}$ is the flux in units of $10^{-9}\, \mathrm{erg\, cm^{-2}\, s^{-1}}$. We use $\eta=0.1$, $f_{\rm ang}=1$ \citep{ibragimov2009accreting}, $k_A = 1$, and D = 10 kpc in further calculations to estimate the magnetic dipole moment. From the joint spectral analysis of XMM-Newton and NuStar observations, $R_{\rm{in}}\sim$12 km and ${F}_{b}\sim 3.23 \times10^{-9}$ erg cm$^{-2}$ s$^{-1}$ in the 0.1-100 keV energy band \citep{malacaria2025disk}. The inferred magnetic dipole moment using these values is $\sim 1.3 \times 10^{26}~\rm{G ~cm^3}$ and the estimated magnetic field ($B = \frac{2 \mu}{R^3}$) is $\sim 2.6\times 10^8$ G. The inferred magnetic field lies within the lower and upper limits ($3.8\times 10^7~\rm{G} < B < 1.6\times 10^9~\rm{G}$) given by \citet{li2025timing}. The estimated mass accretion rate during this period is $\sim 3.26 \times 10^{-9} ~M_{\odot} ~\mathrm{yr}^{-1}$ for ${F}_{b}\sim 3.23 \times10^{-9}$ erg cm$^{-2}$ s$^{-1}$ \citep{malacaria2025disk}. Therefore, using B $\sim 2.6\times 10^8$ G and $\dot{M} \sim 3.26 \times 10^{-9} ~M_{\odot} ~\mathrm{yr}^{-1}$ in equation \ref{eq6}, the ratio $\frac{r_{\rm m}}{R_{*}} \sim$2 (which is within a factor of 2 of the required value for jet formation), which suggests that jet formation may be possible in this system. The main uncertainty in this calculation is the value of the magnetic field strength, which can be as low as $\sim 4\times 10^7$ G for this source \citep{li2025timing}. It should also be noted that the role of the magnetic ﬁeld in jet production and inhibition is still an open question \citep{tetarenko2018radio}. However, the magnetic field estimated from the observed spin-up rate of the pulsar is $\sim 4\times 10^7$ G \citep{li2025timing} and using $\dot{M} \sim 3.26 \times 10^{-9} ~M_{\odot} ~\mathrm{yr}^{-1}$ yields $\frac{r_{\rm m}}{R_{*}} \sim$0.7, which suggests formation of a jet in the system. The discrepancy in $\frac{r_{\rm m}}{R_{*}}$ compared with the value found above is due to the uncertainty in the value of the magnetic field strength. Thus, the radio emission may be attributed to a  compact jet from the accreting pulsar. In the AMXP SAX J1808.4-3658, an X-ray reflare was detected after approximately 15 days of the peak of the main X-ray outburst in 2019. This reflare was associated with the radio brightening of the source and was suggested to be linked to the presence of a jet in this source \citep{gasealahwe20232019}. An X-ray flare and radio detection of SRGA J144459.2-604207 happened within a week of the peak of the main X-ray outburst, which is suggested to be associated with the formation of a jet. However, in SRGA J144459.2-604207, the radio brightening of the source cannot be probed due to a lack of radio monitoring observations.

\section{Conclusions}
We have performed a spectral study of the accreting millisecond pulsar SRGA J144459.2-604207 during the decay of its 2024 outburst, reflares and quiescent state using the NICER and Swift observations. The spectra are fitted using the absorbed Comptonized model with $\Gamma \sim$2. In addition, we also fit the quiescent spectra using the absorbed power-law ($\Gamma \sim$2.2-2.3) and absorbed blackbody model ($kT_{\rm bbody} \sim$0.9 keV). The estimated luminosity in the 0.5-10 keV energy band during the outburst decay is found to be roughly ten times higher than that estimated during reflares. However, the estimated luminosity in the same energy band during the quiescent state is found to be approximately two to five times smaller than that obtained during reflares. We have also explored the spectral evolution of the pulsar during the outburst decay, reflares and quiescent state. The long-term quiescent X-ray activity of the source for approximately 45 years is explored using archival observations, which may be explained using the deep crustal heating model. We find that the quiescent X-ray luminosity of the source cannot be powered alone by the coronal activity of the companion star.
The source is placed on the radio-X-ray luminosity plane, and its position is compared with other accreting sources. We compare the estimated luminosity during reflares and quiescent state during the 2024 outburst with the propeller luminosity of the source and find that the source was accreting above the propeller luminosity. One of the reflares among several reflares detected during the outburst decay was near-simultaneous with the detection of an ultrafast outflow and radio emission. We explore possible mechanisms that may power reflare, outflow, and jet formation in this source. Simultaneous multi-wavelength observations of the accreting pulsar during future outbursts are required to study and constrain the mechanisms of outflows and disc-jet coupling in this pulsar.

\begin{acknowledgments}
We are thankful to the reviewer for carefully
going through the manuscript and making detailed, valuable, and
constructive suggestions that have greatly improved the presentation
of this paper.
This research has made use of software provided by the
High Energy Astrophysics Science Archive Research Center (HEASARC),
which is a service of the Astrophysics Science Division at NASA/GSFC and
the High Energy Astrophysics Division of the Smithsonian Astrophysical
Observatory. NICER is a 0.2-12 keV X-ray telescope operating on
the International Space Station, funded by NASA. We acknowledge the use of Swift public data archive. This research has made use of the MAXI \citep{matsuoka2009maxi} light curve provided by RIKEN, JAXA, and the MAXI team. This research has made
use of the SIMBAD astronomical database \citep{wenger2000simbad} and NASA’s
Astrophysics Data System. ADC acknowledges
support from ARIES through post-doctoral fellowship.
\end{acknowledgments}
\facilities{NICER, Swift (XRT), MAXI}

\software{HEASOFT \citep{heasarc2014heasoft}, FTOOLS \citep{blackburn1995ftools}, XSPEC \citep{arnaud1996astronomical}, Numpy \citep{harris2020array}, Matplotlib \citep{hunter2007matplotlib}
          }


%\appendix

%% For this sample we use BibTeX plus aasjournalv7.bst to generate the

\bibliography{AAS68609}{}
\bibliographystyle{aasjournalv7}



\end{document}